# Single-mode propagation of light in one-dimensional all-dielectric light-guiding systems

Changbiao Wang

ShangGang Group, 70 Huntington Road, Apartment 11, New Haven, CT 06512, USA
changbiao_wang@yahoo.com

Numerical results are presented for single-mode guidance, which is based on photonic band gap (PBG) effect, in one-dimensional planar all-dielectric light-guiding systems. In such systems there may be two kinds of light-speed point (the intersection of a mode-dispersion curve and the light line of guiding region ambient medium): one is the intrinsic light-speed point that is independent of the guiding region width, and the other is the movable light-speed point that varies with the guiding region width. It is found that the intrinsic light-speed point plays an important role to form the single-mode regime by destroying the coexistence of the lowest guided TM and TE modes that are born with a degeneration point. A sufficient and necessary condition for intrinsic light-speed points is given. The transverse resonance condition is derived in the Maxwell optics frame, and it is shown that there is a significant revision to the traditional one in the ray optics model. A mode-lost phenomenon is exposed and this phenomenon suggests a way of how to identify PBG-guided fundamental modes. Quasi-cutoff-free index-guided modes in the PBG guiding structures, which appear when the higher-index layers are adjacent to the guiding region and the guiding region width is small, are exposed and analyzed as well.

Keywords: single-mode guidance, all-dielectric light-guiding systems, PBG-guided

PACS: 41.20.-q, 42.25.Bs, 42.81.Qb

## 1. INTRODUCTION

In all-dielectric light-guiding systems, single-mode propagation of light by photonic band gap (PBG) effect [1,2] has aroused extensive interest [3-9]. Such single-mode guidance was first demonstrated by experiments for two-dimensional (2D) honeycomb PBG fibers [3], and subsequently it was also predicted in an analysis of an all-dielectric coaxial waveguide [4]. Planar periodically stratified uniform medium inserted with an empty channel, called Bragg reflection waveguide [10], is the simplest 1D PBG all-dielectric light-guiding system, and it has been investigated in various approaches [10-17]. In such a symmetric guiding system, however, the possibility of single-mode guidance has never been reported due to the coexistence attribute of guided TM and TE modes [17].

In a typical PBG planar light-guiding system, as shown in Fig. 1, two symmetric bilayer periodical dielectric structures perform as PBG mirrors to confine light in the guiding region that has a lower refractive index [10,11]. The light line of the guiding region ambient medium, which is often used in an analysis of dispersion curves, is defined by $\omega/c = k_z/n_0$ on the $k_z$-$\omega/c$ plane, where $\omega$ (> 0) is the angular frequency, $c$ is the vacuum light speed, $k_z$ is the axial wave number, and $n_0$ is the guiding region refractive index. A light-speed point is defined as the intersection of a mode-dispersion curve and the light line, and at this point the wave phase velocity $v_{ph} = \omega/k_z$ is equal to the light speed $c/n_0$. There may be two kinds of light-speed points in the planar guiding structures. One is termed to be "intrinsic" light-speed point, which does not move on the $k_z$-$\omega/c$ plane when the guiding region width changes; the other is termed to be "movable" light-speed point, which moves when the guiding region width is adjusted.

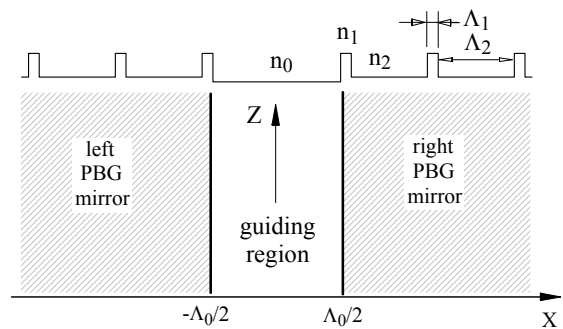

Fig. 1. One-dimensional all-dielectric PBG light-guiding system, consisting of a guiding region and two symmetric half-infinite bilayer periodical structures. $\Lambda_1$ (> 0) and $\Lambda_2$ (> 0) are, respectively, the thickness of the first and second layers; $n_1$ and $n_2$ are their corresponding refractive indices. The structure period is $\Lambda = \Lambda_1 + \Lambda_2$, and the width and refractive index of guiding region are, respectively, $\Lambda_0$ and $n_0$.

A single-mode regime refers to a frequency range within which there is only one non-degeneration guided mode, that is, single-polarization single-mode [18]. According to the band area where the single-mode frequency range is located, the single-mode regime can be divided into two kinds in the planar PBG guiding systems. The first kind of single-mode regime is defined as the single-mode frequency range that is located within the PBG reflector's band or the band part above the light line of the guiding region ambient medium (called "light line" for short in all of what follows unless specified); namely reflector-band single-mode regime [confer Fig. 2(*c*)]. The second kind of single-mode regime is defined as the single-mode frequency range that is extended over the full band, no matter above or below the light line; namely full-band single-mode regime (confer Fig. 5). The reflector-band single-mode regime is a conditional single-mode frequency range, where the other modes out of the reflector's band need to be suppressed by controlling the polarization and/or location of exciting source, while the full-band single-mode regime is a complete single-mode frequency range, where no other modes can be existent in principle. Apparently, the full-band single-mode regime has a narrower frequency range.

A cutoff-frequency difference of lowest two adjacent modes is often used to judge a single-mode regime. Recently, a mode analysis approach based on the transverse resonance condition [17] has been developed and it can be used for examining the mode cutoff frequencies, but it is only applicable to the fast-wave modes wholly above the light line, because all the slow-wave modes below the light line, no matter how close to the light line a slow-wave mode is, are artificially "cut off", although those "cutoff" modes may be existent strongly, and some of those slow-wave modes have been suggested for particle acceleration [19,20]. Accordingly, this approach [17] is not suitable for identifying a real cutoff-frequency difference of two adjacent modes. Moreover, due to the complexity of the distribution of dispersion curves for PGB guiding structures, usually one cannot predict under what conditions the cutoff-frequency difference has a maximum employable frequency range and how well the light field can be confined in the guiding region for this range. To reliably explore the existence of single-mode guidance, it is necessary to examine the evolution of the complete mode structure and the field distributions of potentially usable modes by using a field-solution method.

The planar all-dielectric light-guiding system [10,11] actually is a transversely periodical boundary-value problem in classical electromagnetic theory, where two perfect conducting planes of the parallel-plate waveguide are replaced by two symmetric periodic structures. Usually, PBG-guided TM and TE modes are born in pairs with a degeneration point, and the fundamental modes cannot be separated if the parameters of the periodic structures and guiding region are not properly taken. In this paper, extending our previous work [21], we would like to present numerical results from a field-solution method, showing the existence of single-mode regime in the planar all-dielectric guiding system [10,11].

Unlike 2D and 3D PBG structures where pure numerical simulations are needed, the planar guiding system is easier to handle, and may provide a deep insight of physics. Asymptotic analysis indicates that the large-radius cylindrical Bragg fiber is analogous to a planar structure [19] and a better understanding of the physics in the planar structure has a fundamental importance. In this study, through examining the single-mode guidance we present some new fundamental properties of the planar PBG guiding structure, which include: (1) mode degeneration points, (2) mode-lost phenomenon, and (3) the revised transverse resonance condition in Maxwell optics model. These fundamental properties are helpful in further understanding the physics of PBG guiding structures, and also play a significant role in the understanding of numerical results in a number of previous publications.

The paper is organized as follows. In section 2, classifications of planar light-guiding systems and electromagnetic modes are described. In section 3, numerical results with theoretical analysis are presented, and finally in section 4, some conclusions and remarks are given.

## 2. CLASSIFICATIONS OF PLANAR LIGHT-GUIDING SYSTEMS AND ELECTROMAGNETIC MODES

In the planar guiding systems, there are two types of modes, TM and TE modes, and they only have three non-zero field components: $E_x$, $E_z$, and $H_y$ for TM modes, and $H_x$, $H_z$ and $E_y$ for TE modes. The PBG light-guiding systems guide light at electromagnetic modes in stop bands (band gaps). Similar to the refractive index guidance in conventional optical fibers, the PBG guidance is another form of total internal reflection [21]. General speaking, the mode structure in such a dielectric system is quite complicated. But it is much simplified if we divide all modes into two main groups. One is the fast-wave mode of which dispersion curves lie within the light cone formed by the guiding region ambient medium (confer Fig. 2), and the other is the slow-wave mode of which dispersion curves lie outside the light cone. In the guiding region, fast-wave mode fields are distributed in the form of standing wave in the transverse ($x$-) direction, while slow-wave mode fields vary exponentially. Although the planar light-guiding system may propagate both fast and slow waves, only the fast-wave modes are defined as PBG-guided modes in this paper, because only in such a case all the transverse wave numbers both in the guiding region and periodic structures are real and the periodic structures perform as real mirrors in a usual meaning.

As it is well known, the definition of mode order in regular metallic waveguide systems is based on the sequence of zeros of a proper function or its derivative, depending on the kind of boundary condition [22]. In the planar bilayer guiding systems [11], besides mode types (TM and TE), the mode property depends on the arrangement of the layer's wave impedance $(\mu/\varepsilon)^{1/2}$ in the periodic structures, with $\mu$ and $\varepsilon$ the layer's permeability and permittivity respectively, because the PBG mirrors with different wave impedance arrangements behave themselves with different kinds of boundary conditions. Based on this, the planar guiding systems, where all layers of the periodic structures have the same permeability so that the wave impedance arrangement can be described by their refractive index arrangement, can be divided into two kinds. The first kind of planar guiding system is the one where the refractive index $n_1$ of the first layer (adjacent to the guiding region) is larger than the second one $n_2$ (confer Fig. 1); called first kind of index arrangement. The second kind is the one where $n_1$ is smaller than $n_2$ (confer Fig. 6); called second kind of index arrangement.

For the first kind of planar light-guiding system, the number of zeros of $E_x$ in the guiding region is used to define the mode order for TM modes, while the number of zeros of $\partial H_x / \partial x$ is used to define the mode order for TE modes. For the TM$_{mn}$ mode, with $m$ the mode order and $n$ the stop band order, $E_x$ has $m$ zeros and it has the same parity as $m$; namely, $E_x$ is an even (odd) function of $x$ when $m$ is even (odd). For the TE$_{mn}$ mode, $H_x$ has $m$ extremums (peaks), and it has an opposite parity to $m$; namely, $H_x$ is an odd (even) function of $x$ when $m$ is even (odd).

For the second kind of planar light-guiding system, the number of zeros of $\partial E_x / \partial x$ in the guiding region is defined to be the mode order for TM modes, while the number of zeros of $H_x$ is defined to be the mode order for TE modes. For the TM$_{mn}$ mode, $E_x$ has $m$ extremums in the guiding region and it has a parity opposite to the one of $m$. For the TE$_{mn}$ mode, $H_x$ has $m$ zeros and it has the same parity as $m$.

It should be noted that, between the mode order definition given here and the one given by Li and Chiang [17], there are important differences. (*i*). The definition given here does not have any dependence on the guiding region width and the band locations. Especially, there is no minus mode order assigned, which is more consistent with the traditional custom; the same TM-mode definition is given for all band locations, including above and below the Brewster's line. (*ii*). The mode-order definition for the second kind of planar guiding structures is also assigned.

Brewster's point deformed the band structure and twisted dispersion curves. For the convenience of description, the part of the first TM-stop band between the Brewster's point and the $k_z = 0$ line is called "main first TM stop band" (confer Fig. 5), and the main first TM stop band plus the first TE stop band is called "main first stop band". The main first stop band is the most interesting part and it has a relatively simple mode structure.

In the zero[th] stop band, the transverse wave number of the lower-index layer of the bilayer unit cell is imaginary, leading to traditional total internal reflection. Accordingly, the zero[th] stop band is virtually an index-guiding band, and it has a mode structure which is essentially different from the main first stop band, but, to some extent, similar to the one in the (index-guiding) single channel dielectric waveguide [11,18]. When the higher-index layers are adjacent to the guiding region and the guiding region width is small, the planar PBG guiding system may present complicated "mixing PBG-index-guided" mode structures, with some unfamiliar slow-wave modes appearing in the zero[th] stop band (confer Fig. 2), because in such a case the guiding system looks like, in a macro scale, a large dielectric slab with a refractive index peak at the central part — a macro-effect of refractive-index distribution [18]. When the guiding region width is large enough, almost all slow-wave modes will disappear. An informative numerical example for a pure PBG-guided TE-mode structure is demonstrated in [21], where all modes are fast-wave modes, located above the light line of the guiding region ambient medium, or within the PBG reflector's band.

PBG-guided mode dispersion curves shift down towards the low frequency side when the guiding region width increases [21]. It has been shown analytically that, no matter for a fast- or slow-wave mode, the decaying field in the periodic structure is characterized by a decaying factor multiplied by a periodic function that has double the period of the structure if the band order is odd, or has the same period as the one of the structure if the band order is even [21]. It was also shown by Li and Chiang [17] that, for the fast-wave mode field component, the number of the zeros per period in the periodic structures is equal to the band order.

Suppose that $\phi_L$ and $\phi_R$ are the arguments of the reflection coefficients of the left and right PBG mirrors (see Fig. 1), and $\phi_0 = 2k_x^{(0)}\Lambda_0$ is the round-trip shift phase in the guiding region for a guided TM$_{mn}$ or TE$_{mn}$ mode, where the transverse wave number $k_x^{(0)}$ is real ($>0$) for PBG-guided modes. The round-trip shift phase $\phi_0$ satisfies the inequality $2(m+1)\pi > \phi_0 \geq 2(m-1)\pi$ for $m \geq 1$, or $2\pi > \phi_0 \geq 0$ for $m = 0$, because the phase difference between two adjacent zeros or extremums is $\pi$. Thus the mode order $m$ is closely related to the transverse resonance phase defined by $\phi_L - \phi_R + \phi_0$ in the Maxwell optics model, resulting from the fact that $\phi_L - \phi_R + \phi_0$ satisfies the transverse resonance condition $\phi_L - \phi_R + \phi_0 = m_{TR}(2\pi)$, where the integer $m_{TR}$ can be $(m-1)$, $m$, or $(m+1)$ for $m \geq 1$, and 0 or 1 for $m = 0$, with $\phi_R = -\phi_L$ (see Appendix A). One important difference between the boundary conditions on the metallic mirror and the dielectric PBG mirror is in that the tangential (normal) component of the electric (magnetic) field on the metallic mirror must be equal to zero, while the field components on the dielectric PBG mirror do not have such an independent-of-frequency locked relation with the zero or extremum. In other words, the dielectric PBG mirror is a kind of dispersive reflector. That is why the transverse resonance condition and the mode order are not one-to-one corresponding [21], which also has been confirmed by computations.

## 3. NUMERICAL RESULTS WITH THEORETICAL ANALYSIS

In this section, two kinds of light-speed points in the planar PBG guiding structures are introduced and it will be seen why the lowest two guided TM and TE modes can possess different kinds of light-speed points. By combining the numerical results with theoretical analysis, it will be seen how the lowest two modes can be separated to form a single-mode regime, and it will be also seen why quasi-cutoff-free (index-guided) all-slow-wave modes may appear when the higher-index layers are adjacent to the guiding region. Some other slow-wave modes appearing during the evolution of mode structures are presented as well.

A necessary condition of the existence of significant single-mode guidance regime is that the parameters of periodic structures are taken so that an intrinsic light-speed point can be created on the $TM_{01}$ mode within the main first TM stop band for the first kind of guiding structure (Fig. 1). To this end, the dielectric constants for the bilayer periodic structure in this numerical example are taken to be 21.2 and 2.6, corresponding to the refractive indices $n_1 = 4.6$ and $n_2 = 1.6$ [4]. The guiding region medium is taken to be empty ($n_0 = 1$). In such a case, the $TM_{01}$-mode dispersion curve intersects the light line, as shown in Fig. 2(a). This intersection point is an intrinsic light-speed point and it is independent of the width of the guiding region. Consequently, when the guiding region width increases, the intrinsic light-speed point does not move and the whole $TM_{01}$-mode dispersion curve only can rotate around the intrinsic light-speed point and keeps itself getting close to the light line; at the same time the whole $TE_{01}$-mode curve is moving toward the allowed band, as seen in Fig. 2(b). For a proper guiding region width, about 1.5 times the index period, almost the whole $TE_{01}$-mode curve is "lost" into the allowed band and a frequency range of single-mode guidance in the reflector's band is formed, as seen in Fig. 2(c). However, if the width is too large, new modes are coming out from the top stop band boundary so that the single-mode regime disappears, as seen in Fig. 2(d).

From Fig. 2, two important conclusions can be drawn. (i). Guided TM and TE modes in the main first stop band are born in pairs with an identical frequency at $k_z = 0$ (degeneration point) when the guiding region width increases. The degeneration point comes from the fact that, for a general planar PBG guiding structure no matter whether it is symmetric, TM and TE modes at this point are both corresponding to perpendicularly incident waves (TEM waves with respect to the $x$-propagation direction) and they are completely the same physically, except for a 90-degree difference of the fields in space [21]; in other words, the two perpendicularly-incident TEM waves satisfy the same wave equation and the same boundary conditions, thus leading to the same dispersion equation and, of course, the same eigen frequencies as long as they exist. (ii). By a proper choice of the system's parameters, the evolution of mode-curve distributions under the influence of the intrinsic light-speed point may result in a unique dispersion curve to live in a certain frequency range above the light line, that is, a reflector-band single-mode regime.

One might question: Why is the intrinsic light-speed point required? Computations show that, in the planar guiding structure where the light line is below the Brewster's line, there will be no intrinsic light-speed point within the main first TM stop band, and the $TM_{01}$ mode will be "lost" into the allowed band with the $TE_{01}$ mode together as guiding region width increases, resulting in that always at least one TM mode and one TE mode coexist at the same frequency within the reflector's band. This phenomenon also can be clearly identified from the distributions of TE- and TM-mode dispersion curves shown in the Fig. 7 and Fig. 9 of the recently published work by Li and Chiang [17]. [In both their Figs. 7 and 9, TE and TM modes have the same frequencies or $V$-values at $b = -0.8$ (corresponding to $k_z = 0$). When $b < -0.8$ in their Fig. 7, $k_z^2 < 0$ must hold, leading to an imaginary $k_z$, and no guided TE modes should exist, that is, the parts of curves for $b < -0.8$ are not physical.]

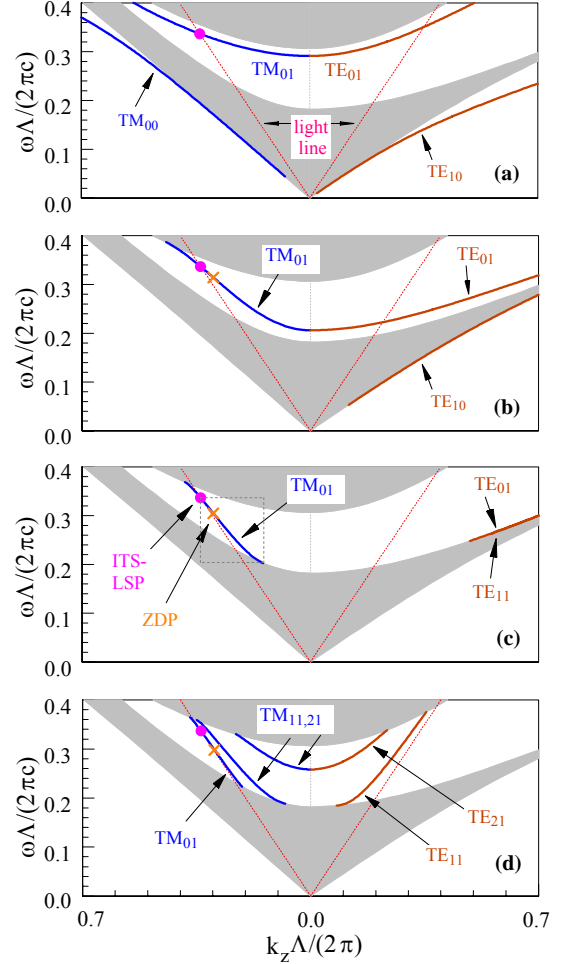

Fig. 2. Evolution of mode structure with increase of the width of guiding region. Left side of the $k_z = 0$ line: TM mode; right side: TE mode; $k_z > 0$ for the both sides. White: stop band; gray: allowed band. Periodic structure's parameters: $n_1 = 4.6$, $n_2 = 1.6$; $\Lambda_1 = 0.3$ μm, $\Lambda_2 = 2.15$ μm; guiding region refractive index: $n_0 = 1$. (a). $\Lambda_0 = 0.02$ μm (0.008 $\Lambda$), (b). $\Lambda_0 = 1$ μm (0.41 $\Lambda$), (c). $\Lambda_0 = 3.6$ μm (1.47 $\Lambda$), and (d). $\Lambda_0 = 9.8$ μm (4 $\Lambda$). The thickness ratio of the two layers is taken so that the intrinsic light-speed point (ITS-LSP, solid circle) is reasonably close to the top boundary of the first stop band to improve the bandwidth of single-mode guidance. The zero-dispersion point (ZDP, cross) shifts towards the low frequency side with the increase of the guiding-region width $\Lambda_0$. The dashed box in (c) shows the reflector-band single-mode regime. Properties of mode structure in the PBG reflector's band of the main first stop band: (i). Only the fundamental $TM_{01}$ and $TE_{01}$ modes exist when $\Lambda_0$ is small as seen in (a) and (b); (ii). Every pair of guided TM and TE modes is born with the same frequency at $k_z = 0$ (degeneration point); (iii). The fundamental $TE_{01}$ mode completely disappears when $\Lambda_0$ is large enough as seen in (d) — called "mode-lost phenomenon"; (iv). All PBG-guided mode curves shift down towards the low-frequency side as $\Lambda_0$ increases.

To get a better understanding of why the intrinsic light-speed point is "intrinsic", let us take a look of its origin and feature. The intrinsic light-speed point originates from the equation $(t_{11}-\sigma)(t_{22}-\sigma)/t_{12}=0$ or $t_{21}=0$, where $t_{11}$, $t_{12}$, $t_{21}$, and $t_{22}$ denote the elements of the real eigenmatrix $T$ for a half-infinite bilayer periodic structure (see Appendix B). It can be analytically shown that for a planar bilayer periodic guiding system, a sufficient and necessary condition for the discrete roots of $t_{21}=0$ on the light line to satisfy their dispersion equation independently of the guiding region width is $|t_{11}|>1$ (Appendix B).

Suppose $\mathrm{Min}(n_1, n_2) > n_0$ holds for the planar guiding structure; that is, the guiding region has a lower refractive index. Inserting the light-line expression $k_z = n_0\omega/c$ into $t_{21}^{(\mathrm{TM})} = 0$ and $t_{21}^{(\mathrm{TE})} = 0$, we obtain

$$\frac{n_{10}}{n_{20}}\frac{\tan(n_{10}k_0\Lambda_1)}{\tan(n_{20}k_0\Lambda_2)} = \begin{cases} -\varepsilon^{(1)}/\varepsilon^{(2)}, & (\text{TM mode}) \\ -\mu^{(1)}/\mu^{(2)}, & (\text{TE mode}) \end{cases}, \quad (1)$$

where $\tan(n_{20}k_0\Lambda_2) \neq 0$, $n_{10} = (n_1^2 - n_0^2)^{1/2}$, $n_{20} = (n_2^2 - n_0^2)^{1/2}$, $k_0 = \omega/c$, and $\varepsilon^{(i)}$ and $\mu^{(i)}$ ($i = 1, 2$) are, respectively, the dielectric permittivity and permeability in the bilayer unit cell. (In the following analysis, $\mu^{(1)} = \mu^{(2)}$ is assumed.) Obviously, the discrete frequency roots of Eq. (1) are independent of the guiding region width. For the first kind of index arrangement, only the TM-mode roots of Eq. (1) can make $|t_{11}|>1$ hold, and these roots automatically satisfy their dispersion equation independently of the guiding region width. Accordingly, all the points ($k_z = n_0\omega/c$, $\omega/c$) in the $k_z$-$\omega/c$ plane, corresponding to the TM-mode discrete frequency roots, are intrinsic light-speed points. To ensure that a root of Eq. (1) is located within the main first TM stop band, the light line must be above the Brewster's line to pass through the main TM stop band (confer Fig. 5). That means that $n_1/n_0 > (1 + n_1^2/n_2^2)^{1/2}$, or $n_1/n_0 > n_2/(n_2^2 - n_0^2)^{1/2}$ must hold, because the light line is given by $k_0/k_z = 1/n_0$ while the Brewster's line [11] is given by $k_0/k_z = (n_1^2 + n_2^2)^{1/2}/(n_1 n_2)$. The intrinsic light-speed point only can be found on the lowest even mode in a stop band, and at this point the electromagnetic field in the guiding region has a distribution with $E_z = 0$ and $H_z = 0$, that is, a TEM distribution. Guided TM$_{01}$ mode is the lowest even mode in the main first TM stop band, and it can have the intrinsic light-speed point.

As mentioned previously, there is another kind of light-speed points that depend on the guiding region width — movable light-speed points; they are obtained by inserting $k_z = n_0\omega/c$ into the dispersion equations and excluding those roots of Eq. (1) (if applicable). At such a point, the field in the guiding region has a TM ($H_z = 0$) or TE ($E_z = 0$) distribution, instead of TEM distribution. The movable light-speed point for each TM- and TE- stop band only can be found on the lowest odd mode. Guided TM$_{11}$ and TE$_{01}$ modes are the lowest odd modes in the main first stop band, and they both can have the movable light-speed point.

It is seen from the above that the intrinsic light-speed point is a common root of $t_{21}=0$ and the dispersion equation, while the movable light-speed point is only the root of the dispersion equation. The condition $t_{21}=0$ requires that the eigenvalue of the periodic structure takes a special form ($\sigma = t_{11}$ or $t_{22}$), while the dispersion equation rules the property of a guided mode; thus the intrinsic light-speed point can be taken as a "resonance" point, at which the guided mode displays an unusual feature.

To sum up, the fundamental TM$_{01}$ (even) and TE$_{01}$ (odd) modes are the most interesting modes. TM$_{01}$ mode can intersect the light line only at the intrinsic light-speed point, while TE$_{01}$ mode can intersect the light line only at the movable light-speed point.

Now we can easily understand how the mode coexistence can be destroyed. At first the TM$_{01}$ and TE$_{01}$ modes are born with a degeneration point and both intersect the light line. The light-speed point on the TM$_{01}$-mode curve is an intrinsic light-speed point, and it does not move when the guiding region width increases, as if the mode curve were nailed up in the main first TM stop band and the intrinsic light-speed point behaved like an "axis of rotation". However, the light-speed point on the TE$_{01}$-mode curve is a movable light-speed point, and the whole TE$_{01}$-mode curve is movable so that it can be "lost" into the allowed band. The TM$_{01}$-mode curve is nailed up while the TE$_{01}$-mode curve is not, resulting in the destruction of the coexistence of TM$_{01}$ and TE$_{01}$ modes when the guiding region width is properly adjusted. Obviously, if the "axis of rotation" (intrinsic light-speed point) were not produced within the main TM stop band, the TM$_{01}$-mode curve would be rotated out of the main TM stop band.

It should be point out that the single TM-mode guidance in the symmetric structure presented here is different from the single TE-mode guidance in an asymmetric slab waveguide structure, where two different PBG mirrors without a common first TM-stop band are employed, no intrinsic light-speed point is required, and no restriction is put on the guiding region width [23]; the two single-mode operations are based on different mechanisms.

In numerical calculations, all dispersion curves were classified with the mode order definitions given in this paper, and the field distributions in periodic structures were confirmed to be consistent with the analytic theory given in [21]. It is seen from Fig. 2(*a*) that there are two all-slow-wave mode dispersion curves (TM$_{00}$ and TE$_{10}$, both are lowest modes and even; no TE$_{00}$ mode) along the top boundary of the zero$^{\text{th}}$ TM-TE stop band, which have never been exposed [10-17]. Their appearance can be taken to be a result from the macro-effect of refractive index distribution, as mentioned previously, when considering that the vacuum guiding region width is so small compared with the two adjacent high-index layers (0.02/0.6) so that they behave as an averaged high-index core macroscopically. This can explain why these modes look like the cutoff-free fundamental modes in the traditional single-channel dielectric waveguide [11,18], and also can explain why they disappear when the guiding-region width is large enough, as seen in Fig. 2(c-d).

As shown in Fig. 2(a), $TM_{01}$- and $TE_{01}$-dispersion curves are divided into two parts by the light line. The parts within the light cone are of fast-wave modes, while the parts out of the light cone are of slow-wave modes. It should be pointed out that the slow-wave mode still agrees with the definitions of mode order and parity. Accordingly, only low-order mode curves ($m = 0$ and 1) may intersect the light line while high-order mode curves ($m \geq 2$), which are of all-fast-wave modes, never intersect the light line or extend out of the light cone, resulting from the fact that the high-order mode fields at least have two zeros or extremums in the guiding region while the slow-wave mode fields only can have one zero or extremum due to the property of hyperbolic functions.

It is seen from Fig. 2(c) that, along the bottom boundary of the first TE stop band, the residual $TE_{01}$ mode looks like overlapping with the newly emerging $TE_{11}$ mode. However the two mode curves never intersect because they have different parities; in other words, there is no degeneration between the modes of the same type. A small increase in the guiding region width has little effect on the existence of the two slow-wave modes, while too large an increase will bring about new fast-wave modes coming out from the top band edge, resulting in a smaller single-mode regime instead. These TE slow-wave modes have different polarizations from the one of $TM_{01}$ mode, and they are far below the light line, with most of their EM energy weakly localized in the periodic structures; thus they cannot be excited by the way that is used for launching the guided $TM_{01}$ mode.

As shown in Fig. 2(c), there is also a zero-dispersion point (group velocity extremum) in the $TM_{01}$ single-mode regime, like in the all-dielectric coaxial waveguide case [4]. The normalized group velocity and phase velocity around the zero-dispersion point is shown in Fig. 3. It is seen that the zero-dispersion point has a maximum group velocity of 0.83 and it has a frequency lower than the one of the light-speed point. From (a)-point to (c)-point, the relative wavelength range is over 20% while the change of group velocity is only about 4%, showing that the single-mode guidance has a wide frequency range.

Computations for different planar guiding structures show that the zero-dispersion point can be at any side of the intrinsic light-speed point, or they can overlap each other (confer Fig. 5), depending on the structure's parameters. When the guiding region width increases, the zero-dispersion point shifts towards the low-frequency side, as seen in Fig. 2, for example. When the zero-dispersion point and intrinsic light-speed point overlap, the $TM_{01}$ mode has a TEM field distribution in the guiding region, without any dispersion of group velocity.

A well-recognized implicit assumption for a loss-free waveguiding system is that the energy velocity of a guided wave is equal to its group velocity. The group velocity is calculated from the dispersion equation that is decided by boundary conditions [22], while the energy velocity is evaluated from EM field distributions [19]; they are both structure parameters. This equivalency of energy velocity and group velocity is confirmed in the planar light-guiding system, also shown in Fig. 3. It should be indicated that, only the whole structure's energy velocity, instead of the unit cell's energy velocity [24], is equal to the group velocity, because the periodicity of the planar light-guiding system is defected by the guiding region. The whole structure's, unit cell's, and guiding region's energy velocities are related through $\beta_{en} = (\beta_{en}^{(cell)} + \beta_{en}^{(guid)}\xi)/(1+\xi)$, where $\xi$ is the ratio of the stored energy in the guiding region to the one in all the two half-infinite periodic structures. The more strongly the EM energy is localized in the guiding region, the larger the ratio $\xi$ is, leading to the structure's energy velocity being closer to the guiding region's energy velocity. (Note: only the whole structure's energy velocity in the above three is physical.)

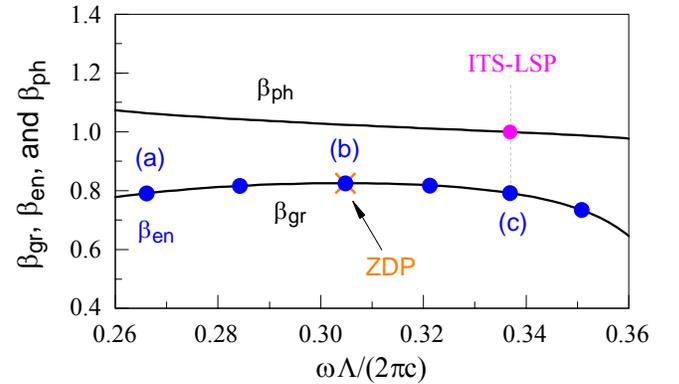

Fig. 3. Normalized-to-vacuum-light-speed group velocity $\beta_{gr}$, energy velocity $\beta_{en}$, and phase velocity $\beta_{ph}$ for the fundamental $TM_{01}$ mode in single-mode regime. Sample $\beta_{en}$-points calculated from light-field distributions excellently fit the $\beta_{gr}$-curve. (a)- and (c)-points have the same group velocity. (b)-point overlaps the zero-dispersion point ZDP.

A single-mode regime without an effective light confinement does not have a practical meaning. Band structure and dispersion-curve distribution only can be used to check whether a single-mode regime is formed; however, the quality of light confinement has to be estimated by examining the field profiles. The field distributions at points (a), (b), and (c) in Fig. 3 are shown in Fig. 4(1). It is seen that the $TM_{01}$-mode light fields are strongly localized in the guiding region, with a TEM distribution at the light-speed point. The transverse resonance phase for the $TM_{01}$ mode around the zero-dispersion point is shown in Fig. 4(2). For this case, the resonance phase satisfies the transverse resonance condition $\phi_L - \phi_R + \phi_0 = 2m_{TR}\pi$ with $m_{TR} = 0$, the same as the mode order of $TM_{01}$. Both the phase shift and the mirror phase compensation are equal to zero after the intrinsic light-speed point, because the $TM_{01}$ mode has become a slow wave (imaginary $k_x^{(0)}$). It also can be seen from Fig. 4(2) that the PBG mirror is of dispersion, unlike a perfect metallic mirror for which the argument of reflection coefficient is a constant.

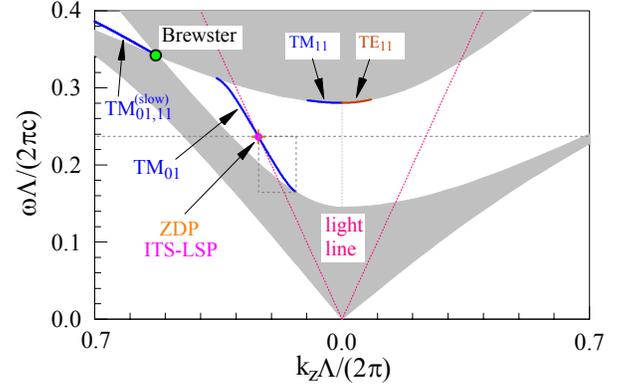

Fig. 5. Mode structure for the planar bilayer PBG guiding system with a full-band single-mode regime. All parameters are the same as the ones in Fig. 2(*c*) except for the bilayer-unit-cell thickness radio $\Lambda_1/\Lambda_2$, which is increased to 0.6/1.85 here. The dashed box shows the full-band single-mode regime, and the intrinsic light-speed point (ITS-LSP) overlaps the zero-dispersion point (ZDP). There are two nearly-overlapping slow-wave modes below the Brewster's line (not drawn), labeled by $TM_{01,11}^{(slow)}$. The main first TM stop band is defined as the band part between the Brewster's point and the $k_z$ = 0 line. It can be intuitively seen that, if the light line were not located above the Brewster's line, then the light line would not pass through the main first TM stop band and there would be no intrinsic light-speed point within the main first TM stop band.

The TEM field distribution in the guiding region is an unusual feature as mentioned before. Mizrahi and Schächter have already shown such a field distribution existing in the planar and coaxial guiding systems that are called Bragg reflection waveguides with a matching layer [25]. For a general guiding system, it is not easy to obtain an analytical solution. Given below is a qualitative analysis of why a TEM field distribution in the guiding region may be supported at discrete frequencies in multi-all-dielectric PBG-mirror guiding systems.

As it is well known from the electromagnetic theory of guided waves, longitudinally translationally invariant multi-conductor microwave guiding systems can support TEM waves while hollow metallic waveguides cannot support any TEM waves [22]. The all-dielectric PBG guiding systems, which can support TM or TE modes while have TEM field distributions in the guiding region at discrete frequencies, can be thought to be evolving from their corresponding multi-conductor microwave guiding systems by replacing all the metallic mirrors with PBG dielectric mirrors. For example, by such a replacement, the 1D planar guiding system [10] can be obtained from a parallel-plate waveguide, while the all-dielectric coaxial waveguide [4] can be obtained from a traditional metallic coaxial waveguide. Because an ideal metallic mirror has no dispersion, the multi-conductor guiding system may support a TEM wave at any frequency (the light line and the TEM-wave dispersion curve overlap), while the PBG mirror is of dispersion (confer Fig. 4) and the PBG guiding system only can support a TEM field distribution in the guiding region at the frequency of the light-speed point. The

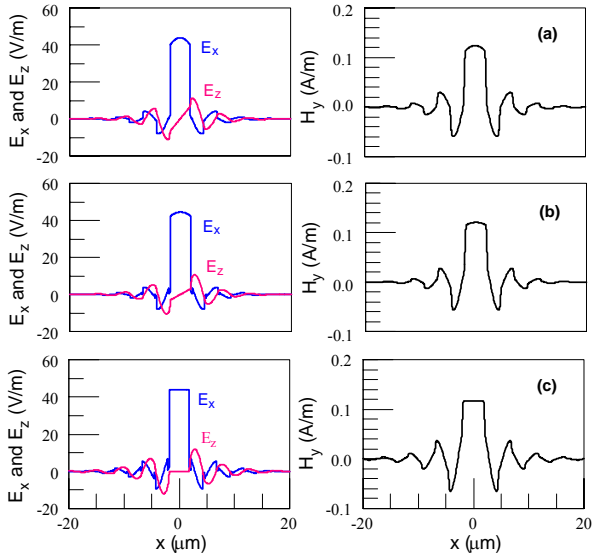

(1). Field distributions.

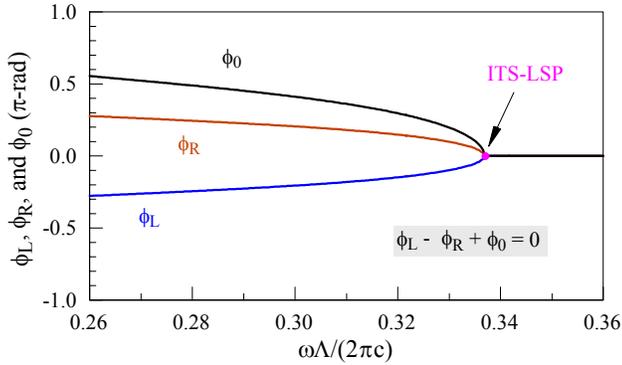

(2). Transverse resonance phase.

Fig. 4. Field distributions and transverse resonance phase. (1). Distributions of electric and magnetic fields at points (*a*), (*b*), and (*c*) in Fig. 3. It is seen from (*c*) that the fundamental $TM_{01}$ mode at the intrinsic light-speed point (ITS-LSP) is a TEM field distribution in the guiding region. (All fields are drawn for a 10 µW/m flowing power.) (2). PBG mirror reflection coefficient arguments $\phi_L$ and $\phi_R$, and round-trip phase shift $\phi_0$ for the $TM_{01}$ mode in single-mode guidance regime. Note: $\phi_L$ and $\phi_R = -\phi_L$ vary with frequency, signifying the dispersion property of PBG mirrors.

As we have seen, the single-mode regime shown in Fig. 2(*c*) is a reflector-band single-mode regime because there are two slow-wave modes left over along the first TE stop band edge. Computations show that, the two slow-wave modes can be further pushed into the allowed band to obtain a full-band single-mode regime by increasing the layer's thickness ratio $\Lambda_1/\Lambda_2$, but leading to the intrinsic light-speed point shifting towards the low frequency side and a reduction in the single-mode frequency range; reducing by 47% when the thickness ratio is increased to 0.6/1.85 from 0.3/2.15, as shown in Fig. 5.

reason why the multi-conductor systems can support a TEM wave results from the fact that the metal wall can provide an axial conduction current $I_c$ so that the Maxwell's equation $\oint \mathbf{H}_\perp \cdot d\mathbf{l} = I_c$ is observed, where $\mathbf{H}_\perp$ is the magnetic field vector of the TEM mode. In the corresponding PBG light-guiding systems, there is no conduction current, but a non-zero displacement electric current $I_{de} = \iint \varepsilon \, \partial E_z / \partial t \, dxdy$ can be supported by the PBG mirror, where $E_z$ is the axial electric field within the periodic structure (PBG mirror) surrounded by all the closed magnetic field lines in the guiding region. Since one of PBG mirrors is required to stand within all the closed field lines in the guiding region in order to support a TEM field distribution, the PBG guiding system must be made up of multi-PBG mirrors. In the first kind of planar guiding system, the fundamental $TM_{01}$ mode at the light-speed point can have such a TEM distribution in the guiding region because $E_x$ of $TM_{01}$ is an even function; for this TEM distribution, all the magnetic field lines in the guiding region will be closed at the infinity.

However, unlike the perfect conductor mirror within which no magnetic field can be supported, a PBG mirror may provide a non-zero displacement magnetic current $I_{dm} = -\iint \mu \, \partial H_z / \partial t \, dxdy$, where $H_z$ is the axial magnetic field within the PBG mirror surrounded by all the closed electric field lines in the guiding region, so that another Maxwell's equation $\oint \mathbf{E}_\perp \cdot d\mathbf{l} = I_{dm}$ is observed. Consequently, another kind of TEM field distribution can be supported by PBG guiding systems. In the second kind of planar guiding system, the fundamental $TE_{01}$ mode at the light-speed point can have such a TEM distribution in the guiding region because $H_x$ of $TE_{01}$ is an even function. Computations show that the second kind of planar guiding system has a smaller mode separation, as shown in Fig. 6, and it is difficult to obtain a significant single-mode regime, which is qualitatively consistent with the results in the all-dielectric coaxial waveguide case [4]. It is also seen from Fig. 6 that the intrinsic light-speed point is on the $TE_{01}$ mode, because in such a case only the TE-mode roots of Eq. (1) satisfy their dispersion equation independently of the guiding region width (see Appendix B).

In the above analysis by comparing the all-dielectric guiding system with the multi-conductor guiding system, an implicit assumption is used, where the magnetic (electric) field line on the PBG mirror surface is parallel to the mirror surface so that any magnetic (electric) field line within the guiding region is closed. It can be generally shown that, if one of two adjacent dielectric zones supports a TEM field distribution while the other supports a TM (TE) field distribution, then the common dielectric interface behaves as an electric (magnetic) wall. Accordingly, this assumption is valid.

For a better understanding of the two kinds of TEM field distributions in the guiding region at intrinsic light-speed points, numerical examples are given in Fig. 7. We can see from Fig. 7 that, both the $TM_{01}$ mode of the first kind of index arrangement and the $TE_{01}$ mode of the second kind have a TEM distribution in the guiding region, respectively; when the index arrangement changes, the mode parity is also changed. It should be point out that the lowest mode ($TE_{01}$) for the first index arrangement, as shown in Fig. 7, is odd, instead of even as defined by Yeh *et al.* [10,11] and West and Helmy [15], where a large guiding region width has already squeezed the real lowest mode out of the first stop band. Note: $H_x$ and $E_y$ have the same parity for TE modes [21].

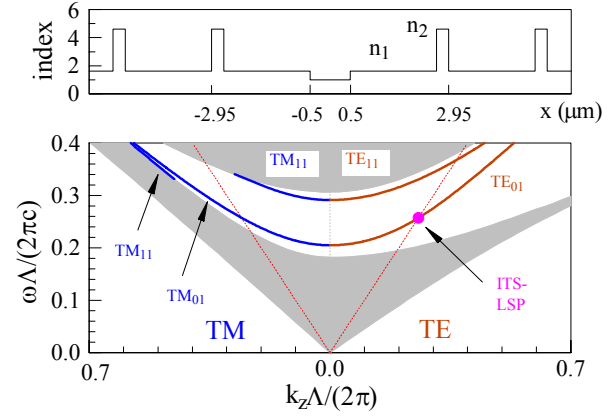

Fig. 6. Mode structure for the second kind of planar light-guiding system that has exactly the same parameters as in Fig. 2(*b*), except for the index arrangement. The $TE_{01}$ mode has an intrinsic light-speed point (ITS-LSP) instead. Compared with Fig. 2(*b*), this guiding system has a smaller mode separation, with $TM_{11}$ (above $TM_{01}$) and $TE_{11}$ (above $TE_{01}$) coming out from the top boundary and closely following $TM_{01}$ and $TE_{01}$. On the bottom boundary, there is an emerging all-slow-wave $TM_{11}$ (below $TM_{01}$) and it will disappear when the guiding region width increase further. Note: the first and second kinds of planar light-guiding systems have the same band structure.

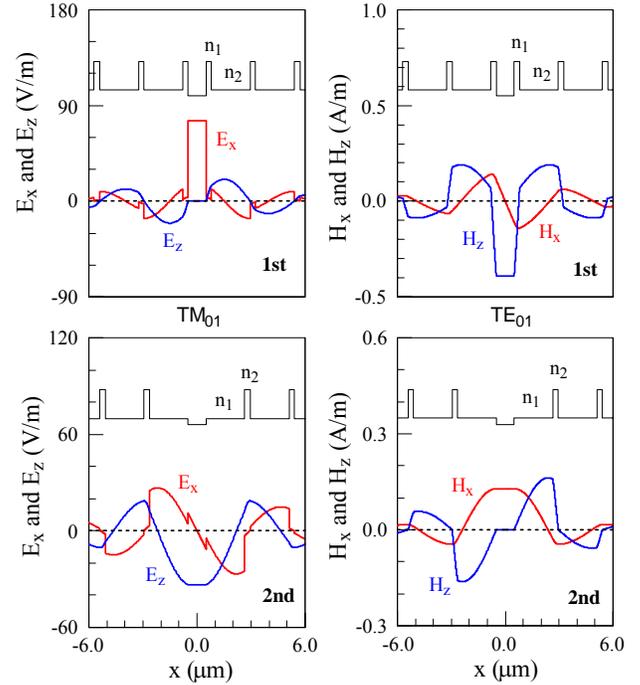

Fig. 7. Field distributions of the fundamental $TM_{01}$ and $TE_{01}$ modes at light-speed points in two different kinds of planar light-guiding systems. The system's parameters are the same as the ones of Fig. 2(*b*) and Fig. 6, respectively. The $TM_{01}$ mode of the 1st kind of guiding system and the $TE_{01}$ mode of the 2nd kind are TEM field distributions in the guiding region. $E_x$-components of the 1st and 2nd kinds have opposite parities, and so do $H_x$-components.

Since the planar all-dielectric light-guiding system [10] and the all-dielectric coaxial waveguide [4] are both dual PBG-mirror guiding systems, they may support a TEM field distribution in the guiding region. In contrast, the hollow-core all-dielectric cylindrical waveguide [26] is a single PBG-mirror guiding system, and it cannot support any TEM field distribution in the guiding region.

## 4. CONCLUSIONS AND REMARKS

By carrying out a detailed and systematic study of TM- and TE-mode structures and field distributions, we have shown the existence of single-mode guidance in the symmetric 1D all-dielectric PBG light-guiding system, that was first proposed and analyzed in the pioneering works [10,11]. This single TM-mode guidance in the symmetric structure presented here is different from the single TE-mode guidance in an asymmetric slab waveguide structure [23]; they are based on different mechanisms. An intrinsic light-speed point is found, which plays an important role in destroying the coexistence of the lowest TM and TE modes that are born with a degeneration point (see Fig. 2). To get such single-mode guidance, the dielectric parameters of periodic structures (PBG mirrors) should be taken so that an intrinsic light-speed point can be created within the main first TM stop band; this conclusion can be used to qualitatively explain why the light line should be above the Brewster's line in the all-dielectric coaxial waveguide case as shown by Ibanescu and coworkers [4], although the light-speed point there is not "intrinsic". Based on a deep analysis of the dispersion equations, we have proved a sufficient and necessary condition for intrinsic light-speed points (Appendix B), which provides strong theoretical support to the numerical results. We also have exposed a macro-effect of refractive index distribution [18] in the planar PBG guiding structures, in which quasi-cutoff-free index-guided modes ($TM_{00}$ and $TE_{10}$ in Fig. 2) may appear in the zero$^{th}$ stop band when the higher-index layers are adjacent to the guiding region and the guiding region width is small.

In the main first stop band, all guided $TM_{01}$ (even), $TM_{11}$ (odd), and $TE_{01}$ (odd) of the first kind of planar guiding structure, and $TE_{01}$ (even), $TE_{11}$ (odd), and $TM_{01}$ (odd) modes of the second kind may intersect the light line by a properly choice of the parameters of periodic structures, but only the $TM_{01}$ (even) mode of the first kind can be used for single-mode guidance. However, $TM_{11}$ (odd) mode of the first kind and $TM_{01}$ (odd) mode of the second kind could be used for particle acceleration [19,20] or the generation of coherent Cherenkov radiation, because they have symmetric axial electric fields and can be slow waves, just like the modes in dielectric-loaded waveguides [27,28]. As shown in Fig. 2 for the first kind of planar guiding structure and Fig. 6 for the second kind, the mode structures include not only TM slow- and fast-wave modes but also TE slow- and fast-wave modes; accordingly, to our best knowledge, they are the most complete out of the mode structures that have ever been reported [17,23,25,29].

It should be pointed out that, in the study of TM modes for control of radiation pressure on mirrors, Mizrahi and Schächter also found the intrinsic light-speed point on a TM mode in their matching-layer Bragg reflection waveguide structure, where the lower-index layers (instead of the higher-index layers) are adjacent to the guiding region [29], while such an intrinsic light-speed point is only on the TE mode in this paper, as shown in Fig. 6. This is because their matching layer is an additional periodicity defect layer, functioning differently; our numerical calculations have confirmed that such a matching-layer structure [29] does not support any significant single-mode regime.

Theoretic analysis often plays a significant role in understanding the results from computations where numerical errors are inevitable and sometimes it is very difficult to judge whether a computational result represents a new physics or is caused by the errors [28]. For example, because of different parities the dispersion curves of two adjacent modes of the same type never intersect theoretically, while they may look like overlapping in computations [confer Fig. 2(*c*) and Fig. 5]. Such a phenomenon also can be identified in the illustrations of dispersion curves for the matching-layer Bragg reflection waveguide structure [29].

Usually there is some form of mode degeneration in a symmetric waveguiding system. For example, in the so-called "single-mode fiber", two orthogonal "polarizations" of the fundamental mode ($HE_{11}$) have a degeneration curve that is cutoff-free [18,30], while in the symmetric single channel dielectric waveguide [11], the TM and TE modes have discrete degeneration points on the light line of cladding medium (corresponding to the critical angle of total internal reflection) [30]. Interestingly, as shown concisely in this paper, even in a general planar guiding system, no matter whether it is symmetric, the PBG-guided TM and TE modes also have degeneration points on the $k_z = 0$ line that corresponds to perpendicular incidence. (The PBG guidance can confine the perpendicular incident wave while the index guidance cannot.) Such degeneration points have been confirmed by computations and were also implicitly included in the illustrations of dispersion curves of the recently published research works [17,23]. It is worthwhile to indicate that, to our best knowledge, it was Li and Chiang who first presented both TM- and TE-mode (fast-wave) dispersion curves at the same time [17], although the planar guiding structure was proposed over thirty years ago [11].

We also have exposed a mode-lost phenomenon, where the fundamental TE and/or TM modes may disappear when the guiding region width is large enough. In a sense, the single-mode regime is realized by taking advantage of the mode-lost phenomenon in such a way that only the fundamental TE mode is allowed to be "lost" while the fundamental TM mode remains, as shown in Fig. 2. Unfortunately, this mode-lost phenomenon has puzzled the community for a long time. For example, in some publications [10,11,15], the defined "lowest mode" turned out to be not really the lowest, while in some others [17], different fundamental-mode definitions were adopted for different guiding region widths. Since the PBG-guided mode dispersion curves shift down towards the low frequency side as the guiding region width increases [21] and the mode-lost phenomenon may appear, it is a good idea to identify and track the lowest modes starting from an enough small guiding region width. Because of the PBG-mirror's dispersion property [confer Fig. 4(2) and Fig. 8], we suggest that the mode-order definition be related to the parity of the lowest mode. If the lowest mode is even (odd), the number of zeros (extremums) in the guiding region is used to define the mode order. However the parity of the lowest mode of the same type may change from one stop band to another. For the sake of simplicity, a unified way of defining TM- or TE-mode order is adopted in this paper, which is based on fitting the PBG-guided modes in the main first stop band.

We have shown in the Maxwell optics frame that, for a symmetric planar guiding structure the two PBG-mirror reflection coefficients are conjugate complex numbers (see Appendix A), while they are taken to be exactly the same in the ray optics model [17]. This unexpected result has significantly revised the traditional transverse resonance condition, which was first introduced by Li and Chiang to study the dispersion relation for fast-wave modes in the planar guiding structure [17]. As shown by theoretical analysis and confirmed by computations in the paper, the mode order and the revised transverse resonance condition are not one-to-one corresponding.

Because the mode properties strongly depend on the distributions of refractive index, the classifications of the planar bilayer PBG guiding systems and the definitions of the electromagnetic modes presented in this paper may not be suitable for other bilayer PBG guiding structures, such as the asymmetric slab waveguide structure where two different PBG reflectors (mirrors) are employed [23], and the matching-layer Bragg reflection waveguide structure where the matching layer, inserted between the guiding region and a quarter-wave stack periodic structure, is usually an additional periodicity defect layer for providing an extra degree of freedom to operate some specified modes at a designed frequency and phase velocity [25].

## APPENDIX A: TRANSVERSE RESONANCE CONDITION

In this appendix, the transverse resonance condition is derived and analyzed in the Maxwell optics frame, where all properties of light are governed by Maxwell's equations. Suppose that the $x$-component of the electric fields for a pure TM-mode standing wave in the guiding region can be written as [21]

$$E(x) = a_0 e^{-ik_x^{(0)}x} + b_0 e^{+ik_x^{(0)}x}$$
$$= 2|a_0| e^{+i\frac{1}{2}(\theta_b + \theta_a)} \cos\left\{ \begin{array}{l} k_x^{(0)}(x - x_c) \\ + \frac{1}{2}[2k_x^{(0)}x_c + (\theta_b - \theta_a)] \end{array} \right\}, \quad (A1)$$

where $a_0 = |a_0|e^{i\theta_a}$ and $b_0 = |b_0|e^{i\theta_b}$ with $|b_0| = |a_0|$, and $x_c = (x_L + x_R)/2$ is the guiding region center with $x_L$ and $x_R$, respectively, the left and right PBG-mirror locations.

The general reflection coefficient at any $x$ in the guiding region is defined as

$$\Gamma(x) = b_0 e^{ik_x^{(0)}x} / a_0 e^{-ik_x^{(0)}x} = \cos\phi(x) + i\sin\phi(x), \quad (A2)$$

with its argument given by

$$\phi(x) = 2k_x^{(0)}x + (\theta_b - \theta_a). \quad (A3)$$

Inserting $x = x_L$ for the left mirror and $x = x_R$ for the right mirror into Eq. (A3), we obtain the arguments $\phi_L$ and $\phi_R$.

If the structure is symmetric with respect to $x_c$, from Eq. (A1) we have $2k_x^{(0)}x_c + (\theta_b - \theta_a) = 2s\pi$ for even distributions, and $2k_x^{(0)}x_c + (\theta_b - \theta_a) = (2s+1)\pi$ for odd distributions, with $s$ an integer. Without loss of generality, we suppose that $(i)$. $2k_x^{(0)}x_c + (\theta_b - \theta_a) = 0$ for even modes, leading to $b_0/a_0 = \exp[-ik_x^{(0)}(x_L + x_R)]$, and $(ii)$. $2k_x^{(0)}x_c + (\theta_b - \theta_a) = \pi$ for odd modes, leading to $b_0/a_0 = -\exp[-ik_x^{(0)}(x_L + x_R)]$. For the even-mode case with $(\theta_b - \theta_a) = -k_x^{(0)}(x_L + x_R)$, we have $\phi_L = -k_x^{(0)}\Lambda_0$ and $\phi_R = +k_x^{(0)}\Lambda_0$ where $\Lambda_0 = (x_R - x_L)$ is the guiding region width. For the odd-mode case with $(\theta_b - \theta_a) = -k_x^{(0)}(x_L + x_R) + \pi$, we have $\phi_L = -k_x^{(0)}\Lambda_0 + \pi$ and $\phi_R = +k_x^{(0)}\Lambda_0 + \pi$. No matter for even or odd modes, the two mirror reflection coefficients are conjugate, that is, $\Gamma(x_L) = \Gamma(x_R)^*$. In computations, however, the arguments $\phi_L$ and $\phi_R$ are obtained from the expression $\Gamma(x_{L,R}) = \cos\phi_{L,R} + i\sin\phi_{L,R} = \cos(k_x^{(0)}\Lambda_0) \mp i\sin(k_x^{(0)}\Lambda_0)$ for the even mode case, and $-\cos(k_x^{(0)}\Lambda_0) \pm i\sin(k_x^{(0)}\Lambda_0)$ for the odd mode case, by calling an intrinsic function, DATAN2$(y,x)$ in Fortran math library for example, which only gives the values from $-\pi$ to $\pi$. Since the above two pairs of conjugate complex numbers are symmetric with respect to the real axis and the intrinsic function DATAN2$(y,-1)$ is not continuous at $y = 0$, we have $\phi_L = -\phi_R$ (see Fig. 4 and Fig. 8) except for $\Gamma(x_{L,R}) = -1$ that leads to $\phi_L = \phi_R = $ DATAN2$(0,-1) = \pi$. (If $\Lambda_0 \neq 0$, $\Gamma(x_{L,R}) = -1$ may take place only at discrete frequencies.) Since $\phi_L = \phi_R = \pi$ and $(\phi_L = \pi, \phi_R = -\pi)$ are equivalent in math, theoretically (not in computations) $\phi_L = -\phi_R$ always holds in the sense of ignoring a difference of an integer time $2\pi$. It should be emphasized that this conclusion is independent of the choice of a coordinate system and it is surprisingly different from the traditional idea in the ray optics that the two arguments should be exactly the same (including the sign) due to the symmetry [17].

Since the round trip shift phase is defined by $\phi_0 = 2k_x^{(0)}(x_R - x_L)$ while the left- and right-mirror arguments are given by $\phi_{L,R} = 2k_x^{(0)} x_{L,R} + (\theta_b - \theta_a)$ from Eq. (A3), we find that the original transverse resonance condition $\phi_L - \phi_R + \phi_0 = 0$ is a natural result of Maxwell's equations, no matter whether the system is symmetric. However the rule of counting $\phi_L$ and $\phi_R$ in practical calculations makes it equal to an integer time $2\pi$, that is, $\phi_L - \phi_R + \phi_0 = m_{TR}(2\pi)$, which can be taken to be a generalization of the dispersion equation of a parallel-plate waveguide [31]. It is seen that this transverse resonance condition shows a revision to the traditional one in the ray-optics model [23]. When the PBG structure is symmetric, leading to $\phi_R = -\phi_L$ as mentioned above, the transverse resonance condition can be written as $2\phi_L + \phi_0 = m_{TR}(2\pi)$ or $-2\phi_R + \phi_0 = m_{TR}(2\pi)$, another revised version of the one in [17].

As indicated in the previous work [21] and as shown by theoretical analysis in this work, the transverse resonance condition and the mode order are not one-to-one corresponding. Such a supporting numerical example is shown in Fig. 8, where the $TE_{31}$-mode dispersion curve was given in [21]. From Fig. 8, it is seen that $\phi_0$ on the whole $TE_{31}$-mode curve changes from $6.38\pi$ to $5.94\pi$. Because the change range in $\phi_0$ is small, the guiding region always contains three extremums (confer Fig. 2 of [21]).

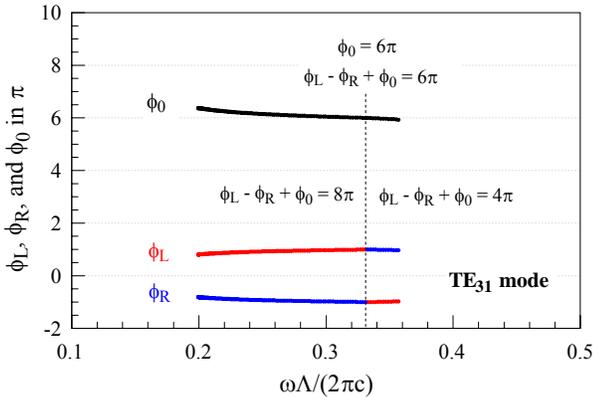

Fig. 8. Profiles of $\phi_L$, $\phi_R$, and $\phi_0$, and the transverse resonance conditions for $TE_{31}$ mode with the mode order $m = 3$ and the band order $n = 1$ [21]. $\phi_0$ changes from $6.38\pi$ to $5.94\pi$ through $6\pi$. $\phi_L = -\phi_R$ except for at $\phi_0 = 6\pi$. At $\phi_0 = 6\pi$ (dashed line), $\phi_L = \phi_R = \pi$, resulting from the discontinuity of the Fortran intrinsic function DATAN2$(y,-1)$ at $y = 0$. After the dashed line, $\phi_L$ and $\phi_R$ exchange the sign. Note: (i). On the left of the dashed line, $\phi_L - \phi_R + \phi_0 = 8\pi$ with $m_{TR} = 4$; (ii). At $\phi_0 = 6\pi$, $\phi_L - \phi_R + \phi_0 = 6\pi$ with $m_{TR} = 3$; (iii). On the right of the dashed line, $\phi_L - \phi_R + \phi_0 = 4\pi$ with $m_{TR} = 2$. The mode order $m = 3$ corresponds to three possible transverse resonance conditions with $m_{TR} = 4$, 3, or 2.

## APPENDIX B: SUFFICIENT AND NECESSARY CONDITIONS FOR INTRINSIC LIGHT-SPEED POINTS

In this Appendix, criterions for intrinsic light-speed points are presented for a symmetric planar bilayer periodic guiding system (confer Fig. 1). Regardless of even or odd modes [10,11,15], the TM- and TE-mode dispersion equations in such a system can be uniformly written in the real form of $D(\omega/c, k_z) \equiv p_{11}p_{22} - p_{12}p_{21} = 0$, where

$$p_{11} = \cos\phi_0 - \frac{(t_{11} - \sigma_L)}{t_{12}} \frac{\chi^{(0)}}{k_x^{(0)}} \sin\phi_0, \quad (B1)$$

$$p_{12} = -t_{12}\cos\phi_1 - (t_{11} - \sigma_R)\frac{\chi^{(1)}}{k_x^{(1)}}\sin\phi_1, \quad (B2)$$

$$p_{21} = -\frac{k_x^{(0)}}{\chi^{(0)}}\sin\phi_0 - \frac{(t_{11} - \sigma_L)}{t_{12}}\cos\phi_0, \quad (B3)$$

$$p_{22} = -t_{12}\frac{k_x^{(1)}}{\chi^{(1)}}\sin\phi_1 + (t_{11} - \sigma_R)\cos\phi_1, \quad (B4)$$

with $\chi = \varepsilon$ for TM mode and $\chi = \mu$ for TE mode, the transverse wave number $k_x^{(i)} = (\omega^2 \varepsilon^{(i)} \mu^{(i)} - k_z^2)^{1/2}$, and $\phi_i = k_x^{(i)} \Lambda_i$, $(i = 0, 1, 2)$. The eigenvalue appearing in above Eqs.(B1)-(B4), satisfying $(t_{11} - \sigma)(t_{22} - \sigma) - t_{12}t_{21} = 0$, is chosen by following the physical condition $|\sigma_L| > 1$ for the left mirror and $\sigma_R = \sigma_L^{-1}$ for the right mirror [11], and the eigenmatrix elements $t_{ij}$ $(i, j = 1, 2)$ are given by

$$t_{11} = \cos\phi_1 \cos\phi_2 - \frac{\chi^{(1)}}{k_x^{(1)}}\frac{k_x^{(2)}}{\chi^{(2)}}\sin\phi_1 \sin\phi_2, \quad (B5)$$

$$t_{12} = \frac{\chi^{(1)}}{k_x^{(1)}}\sin\phi_1 \cos\phi_2 + \frac{\chi^{(2)}}{k_x^{(2)}}\cos\phi_1 \sin\phi_2, \quad (B6)$$

$$t_{21} = -\frac{k_x^{(1)}}{\chi^{(1)}}\sin\phi_1 \cos\phi_2 - \frac{k_x^{(2)}}{\chi^{(2)}}\cos\phi_1 \sin\phi_2, \quad (B7)$$

$$t_{22} = \cos\phi_1 \cos\phi_2 - \frac{k_x^{(1)}}{\chi^{(1)}}\frac{\chi^{(2)}}{k_x^{(2)}}\sin\phi_1 \sin\phi_2. \quad (B8)$$

It can be analytically shown [32] that, for the planar bilayer periodic guiding system, a sufficient and necessary condition for the discrete roots of $t_{21} = 0$ to be intrinsic light-speed points is $|t_{11}| = [\cos^2(n_{20}k_0\Lambda_2) + \eta\sin^2(n_{20}k_0\Lambda_2)]^{1/2} > 1$, where $\eta = (\varepsilon^{(1)}/\varepsilon^{(2)})^2\tau$ for TM mode, and $\eta = (\mu^{(1)}/\mu^{(2)})^2\tau$ for TE mode, with $\tau = n_{20}^2/n_{10}^2$. Furthermore, the electromagnetic fields at the intrinsic light-speed point have a TEM distribution in the guiding region. One might question: Are all intrinsic light-speed points, that are existent, included in the roots of $t_{21} = 0$? The answer is yes [32]. However, not necessarily all the roots of $t_{21} = 0$ can be intrinsic light-speed points; those roots, which also satisfy $\sin(n_{20}k_0\Lambda_2) = 0$ or $\tan(n_{20}k_0\Lambda_2) = 0$ so that $|t_{11}| > 1$ cannot be fulfilled, are not intrinsic light-speed points. Usually such situations take place for some particular bilayer guiding structures; a quarter-wave stack guiding structure, for example.

What is happening if $n_1 = n_0$ or $n_2 = n_0$ (one of the layers of the unit cell has the same refractive index as the guiding region) [17]? When $n_1 = n_0$ or $n_2 = n_0$, we have $\eta\sin^2(n_{20}k_0\Lambda_2) = 0$;

consequently, $|t_{11}|>1$ cannot be fulfilled, and there is no intrinsic light-speed point in such cases. In the following, with additional assumptions of $\text{Min}(n_1,n_2)>n_0$ and $\mu^{(1)}=\mu^{(2)}$ imposed, an equivalent criterion to judge the intrinsic light-speed point is given, which can be directly derived from the above one. This criterion seems to be more convenient for practical applications.

Suppose that $\text{Min}(n_1,n_2)>n_0$ and $\mu^{(1)}=\mu^{(2)}$ hold in a planar bilayer periodic guiding system. This criterion states:

(*i*). For TM mode, a sufficient and necessary condition for the roots of $t_{21}^{(\text{TM})}=0$ with $\tan(n_{20}k_0\Lambda_2)\neq 0$ to be intrinsic light-speed points is

$$n_1 > \text{Max}\left(n_2, \frac{n_2 n_0}{\sqrt{n_2^2-n_0^2}}\right), \tag{B9}$$

or

$$n_1 < \text{Min}\left(n_2, \frac{n_2 n_0}{\sqrt{n_2^2-n_0^2}}\right), \tag{B10}$$

(*ii*). For TE mode, a sufficient and necessary condition for the roots of $t_{21}^{(\text{TE})}=0$ with $\tan(n_{20}k_0\Lambda_2)\neq 0$ to be intrinsic light-speed points is

$$n_1 < n_2. \tag{B11}$$

For the first kind of index arrangement ($n_1>n_2$), only Eq. (B9) can be fulfilled ($n_1=4.6$, $n_2=1.6$, and $n_0=1$, for example), and only TM-mode intrinsic light-speed points may exist, and further more, they must be within the main TM stop bands (as shown in Fig. 2), because Eq. (B9) requires that $n_1/n_0 > n_2/(n_2^2-n_0^2)^{1/2}$ hold, that is, the light line be above the Brewster's line. Inversely, if the light-line is below the Brewster's line ($n_1=1.45$, $n_2=1.3$, and $n_0=1$, for example), Eq. (B9) cannot be fulfilled and there is no any intrinsic light-speed point to be produced.

For the second kind of index arrangement ($n_1<n_2$), Eq. (B11) is directly satisfied and the roots of $t_{21}^{(\text{TE})}=0$ with $\tan(n_{20}k_0\Lambda_2)\neq 0$ are TE-mode intrinsic light-speed points. But, whether Eq. (B10) is fulfilled depends on the choice of dielectric parameters. For example, when $n_1=1.6$, $n_2=4.6$, and $n_0=1$, Eq. (B10) cannot be fulfilled although Eq. (B11) is satisfied, and there is no TM-mode intrinsic light-speed point, as shown in Fig. 6. But Eq. (B10) and Eq. (B11) can be fulfilled at the same time by a proper choice of parameters ($n_1=1.3$, $n_2=1.45$, and $n_0=1$, for example) so that both TE- and TM-mode intrinsic light-speed points can exist; however, in such a case, due to the requirement of $n_1/n_0 < n_2/(n_2^2-n_0^2)^{1/2}$ by Eq. (B10), the light line is below the Brewster's line and the intrinsic light-speed point in the TM stop band is located out of the main TM stop band.

In conclusion, whether the intrinsic light-speed point can exist and what kind of intrinsic light-speed point may exist, depend on the choice of dielectric parameters in the periodic structure and the guiding region.

## References

1. Yablonovitch, E., "Inhibited spontaneous emission in solid-state physics and electronics," Physical Review Letters, Vol. 58, No. 20, 2059-2062, 1987.
2. John, S., "Strong localization of photons in certain disordered dielectric superlattices," Physical Review Letters, Vol. 58, No. 23, 2486-2489 (1987).
3. Cregan, R. F., B. J. Mangan, J. C. Knight, T. A. Birks, P. St. J. Russell, P. J. Roberts, D. C. Allan, "Single-mode photonic band gap guidance of light in air," Science, Vol. 285, No. 5433, 1537-1539, 1999.
4. Ibanescu, M., Y. Fink, S. Fan, E. L. Thomas, and J. D. Joannopoulos, "An all dielectric coaxial waveguide," Science, Vol. 289, No. 5478, 415-419, 2000.
5. Ouyang, G., Y. Xu, and A. Yariv, "Comparative study of air-core and coaxial Bragg fibers: single-mode transmission and dispersion characteristics," Optics Express, Vol. 9, No. 13, 733-747, 2001.
6. Argyros, A., N. Issa, I. Bassett, and M. A. van Eijkelenborg, "Microstructured optical fiber for single-polarization air guidance," Optics Letters, Vol. 29, No. 1, 20-22, 2004.
7. Shephard, J. D., W. N. MacPherson, R. R. J. Maier, J. D. C. Jones, D. P. Hand, M. Mohebbi, A. K. George, P. J. Roberts, and J. C. Knight, "Single-mode mid-IR guidance in a hollow-core photonic crystal fiber," Optics Express, Vol. 13, No. 18, 7139-7144, 2005.
8. Murao, T., K. Saitoh, and M. Koshiba, "Realization of Single-Moded Broadband Air-Guiding Photonic Bandgap Fibers," IEEE photonics Technology Letters, Vol. 18, No. 15, 1666-1668, 2006.
9. Petrovich, M. N., F. Poletti, A. van Brakel, and D. J. Richardson, "Robustly single mode hollow core photonic bandgap fiber," Optics Express, Vol. 16, No. 6, 4337-4346, 2008.
10. Yeh, P. and A. Yariv, "Bragg reflection waveguides," Optics Communications, Vol. 19, No. 3, 427-430, 1976.
11. Yeh, P., A. Yariv, and Chi-Shain Hong, "Electromagnetic propagation in periodic stratified media. I. General theory," Journal of the Optical Society of America, Vol. 67, No. 4, 423-438, 1977.
12. Cho, A. Y., A. Yariv, and P. Yeh, "Observation of confined propagation in Bragg waveguides," Applied Physics Letters, Vol. 30, No. 9, 471-472, 1977.
13. Lekner, J., "Light in periodically stratified media," Journal of the Optical Society of America A, Vol. 11, No. 11, 2892-2899, 1994.
14. Abolghasem, P. and A. S. Helmy, "Matching Layers in Bragg Reflection Waveguides for Enhanced Nonlinear Interaction," IEEE J. Quantum Electronics, Vol. 45, No. 6, 646-653, 2009.
15. West, B. R. and A. S. Helmy, "Properties of the quarterwave Bragg reflection waveguide: theory," Journal of the Optical Society of America B, Vol. 23, No. 6, 1207–1220, 2006.
16. Nistad, B., M. W. Haakestad, and J. Skaar, "Dispersion properties of planar Bragg waveguides," Optics Communications, Vol. 265, No. 1, 153–160, 2006.
17. Li, J. and K. S. Chiang, "Guided modes of one-dimensional photonic bandgap waveguides," Journal of the Optical Society of America B, Vol. 24, No. 8, 1942–1950, 2007.
18. Lee, K. K. Y., Y. Avniel, and S. G. Johnson, "Design strategies and rigorous conditions for single-polarization single-mode waveguides", Optics Express, Vol. 16, No. 19, 15170-15184, 2008.

19. Mizrahi, A. and L. Schächter, "Optical Bragg accelerators," Physical Review E, Vol. 70, No. 1, 016505(1)-(21), 2004.
20. Zhang, Z., S. G. Tantawi, and R. D. Ruth, "Distributed grating-assisted coupler for optical all-dielectric electron accelerator," Physical Review Special Topics — Accelerators and Beams, Vol. 8, No. 7, 071302(1)-(8), 2005.
21. Wang, C., "Light field distributions in one-dimensional photonic crystal fibers," Journal of the Optical Society of America B, Vol. 26, No. 4, 603–609, 2009.
22. Collin, R. E., *Field Theory of Guided Waves*, second ed., IEEE, New York, 1991.
23. Li, J. and K. S. Chiang, "Light guidance in a photonic bandgap slab waveguide consisting of two different Bragg reflectors," Optics Communications, Vol. 281, 5797–5803, 2008.
24. Yeh, P., "Electromagnetic propagation in birefringent layered media," Journal of the Optical Society of America, Vol. 69, No. 5, 742-756, 1979.
25. Mizrahi, A. and L. Schächter, "Bragg reflection waveguides with a matching layer," Optics Express, Vol. 12, No. 14, 3156-3170, 2004.
26. Yeh, P., A. Yariv, and E. Marom, "Theory of Bragg fiber," Journal of the Optical Society of America, Vol. 68, No. 9, 1196-1201, 1978.
27. Wang, C. and J. L. Hirshfield, "Theory for wakefields in a multizone dielectric lined waveguide," Physical Review Special Topics — Accelerators and Beams, Vol. 9, 031301(1)-(18), 2006.
28. Wang, C., "Simulation analysis of rectangular dielectric-loaded traveling wave amplifier for THz sources," Physical Review Special Topics — Accelerators and Beams, Vol. 12, 120701(1)-(15), 2007.
29. Mizrahi, A. and L. Schächter, "Mirror manipulation by attractive and repulsive forces of guided waves," Optics Express, Vol. 13, No. 24, 9804-9811, 2005.
30. Okamoto, K., *Fundamentals of Optical Waveguides*, second ed., Academic Press, Elsevier, 2006; p.76 and p. 82 for $HE_{11}$ mode; Fig. 2.9 on p. 26 for degeneration points.
31. Kong, J. A., "Theory of electromagnetic waves," John Wiley & Sons, New York, 1975.
32. Wang, C., "A sufficient and necessary condition for intrinsic light-speed points in planar all-dielectric guiding systems," ShangGang Rep. 03-2009.